\documentclass[12pt,preprint]{aastex}
\shorttitle{The abnormally hot chromosphere and corona}
\shortauthors{Chen  et al.}
\usepackage[]{lineno}

\begin{document}
%\linenumbers %

\title{The Long-term Evolution of the Solar Transition Region}

\author{W. Q. Chen\altaffilmark{1, 2, 3}, K. J. $Li$\altaffilmark{1,  4}, J. C.  $Xu$\altaffilmark{1, 3}
}
\affil{$^{1}$Yunnan Observatories, Chinese Academy of Sciences, Kunming 650011, China }
%\email{lkj@ynao.ac.cn}}
\affil{$^{2}$University of Chinese Academy of Sciences, Beijing 100049, China }
\affil{$^{3}$Yunnan Key Laboratory of Solar Physics and Space Science, 650216, China }
\affil{$^{4}$State Key Laboratory of Space Weather, Chinese Academy of Sciences, Beijing 100190, China}
%\affil{$^{4}$School of Mathematics and Computer Science, Yunnan Minzu University, Kunming 650504, China}
\email{KJLI:lkj@ynao.ac.cn}

\begin{abstract}
Long-term evolution characteristics of the solar transition region have been unclear.
In this study, daily images of the solar full disk derived from the observations by the Solar Dynamics Observatory/Atmospheric Imaging
Assembly at 304 $\AA$  wavelength from 2011 January 1 to 2022 December 31 are used to investigate long-term evolution of the solar transition region. It is found that long-term variation in the transition region of the full disk  is in phase with the solar activity cycle, and thus the polar brightening should occur in the maximum epoch of the solar cycle.
Long-term variation of the background transition region is found to be likely  in anti-phase with the solar activity cycle at middle and low latitudes.
The entire transition region,
especially the active transition region is inferred to be mainly heated by the active-region magnetic fields  and the ephemeral-region magnetic fields, while the quieter transition region is believed to be mainly heated by network magnetic fields.
Long-term evolution characteristics of various types of the magnetic fields  at the solar surface are highly consistent with  these findings, and thus provide an explanation for them.
\end{abstract}

\section{Introduction}
The anomalous heating of the solar upper atmosphere (the chromosphere, the transition region, and the corona) is an unresolved problem in  solar physics (Cranmer 2012; Kerr 2012; De Moortel $\&$ Browning 2015; De Pontieu et al. 2017; Zirker $\&$ Engvold 2017; Morgan $\&$ Hutton 2018; Li et al. 2018, 2019, 2022; Samanta et al. 2019; Judge 2021).
So far, research on this problem has mainly focused on why high temperatures abnormally occur at a certain location on the solar surface at a particular point in time. However, the problem doesn't just involve abnormally higher temperature than the underneath photosphere, it actually requires an answer to why  the solar upper atmosphere remains abnormally high temperature on the FULL DISK for a LONG TIME.  This means that previous studies are special cases in terms of time and space.
Therefore, a more appropriate way to solve this problem is to analyze the long-term behavior of  magnetic activities on the full solar disk, due to that the solar magnetic fields are the only known source capable of supplying the energy needed for heating the upper atmosphere (Li et al. 2022).

There are four types of the magnetic field on the solar surface: the intra-network magnetic field, the network magnetic field, the ephemeral-region magnetic field, and the active-region magnetic field. The intra-network magnetic fields, the smallest observable magnetic structures at present, may randomly appear anywhere and anytime on the surface of the sun, and the long-term evolution of their occurrence frequency  seems to be unrelated to the solar activity cycle. The network magnetic field is distributed at the full solar disk, and the long-term evolution of its occurrence frequency is in anti-phase with the solar activity cycle (Jin et al. 2011;  Jin $\&$ Wang 2012; Li et al. 2022). The ephemeral-region magnetic field at low latitudes, similar to the magnetic field in active regions, exhibits a butterfly-shaped distribution and is in phase with the solar activity cycle (Jin et al. 2011;  Jin $\&$ Wang 2012, 2015; Li et al 2022). At high latitudes, the ephemeral-region magnetic field undergos long-term evolution along the extended cycle of solar activity (Harvey 1992; Li et al. 2008).
These unique spatiotemporal distribution characteristics of different types of the solar magnetic fields  inevitably delineate the spatiotemporal characteristics of the abnormally heated upper atmosphere. Recently, the long-term evolution of the full-disk quiet chromosphere and that of the full-disk quiet corona are found to be both in anti-phase  with the solar activity cycle,  which are inferred to be caused by the network magnetic fields  (Li $\&$ Feng 2022a, 2022b; Li et al. 2022).  Both the butterfly diagram of the active chromosphere hotter than the background chromosphere and the butterfly diagram of the active corona hotter than the background corona are found to be located directly  above the butterfly diagram of sunspots  colder than the background photosphere, and thus the so-called ``butterfly body" appears throughout the atmosphere. Therefore, the ephemeral-region magnetic field and the active-region magnetic field are inferred to heat the active chromosphere and the active corona, and the butterfly diagram is their common spatiotemporal feature (Li et al. 2022, 2024).

The solar transition region is a relatively special layer in the solar atmosphere, where temperature rises abnormally from the chromosphere outward sharply, while density decreases sharply. Do these aforementioned relations between magnetism and heating in the chromosphere and the corona also apply to the transition region? In this study, we will address this issue.

Polar brightening was first discovered in the photosphere and the chromosphere, which means that the brightness of the photosphere and that of the chromosphere are higher in the minimum time of a solar cycle than in the maximum time (Weber 1865; Sheeley 1964; Li et al. 2022, Sheeley et al. 2011; Chatterjee et al. 2019; Li et al. 2022). However in the corona, polar brightening was found to occur in the corona in the maximum time of a solar cycle, that is, the brightness of the corona is higher in the maximum time than in the minimum time of a solar cycle (Kim et al. 2017; Fujiki et al. 2019; Li et al. 2022; Kumar et al. 2024), which is temporally staggered with the polar brightening of the chromosphere and that of the photosphere, and thus a term  ``polar layered temporally-staggered brightening" is proposed to describe such a phenomenon of the polar brightening in  the entire solar atmosphere. So far, we do not know whether the brightness of the transition region is brighter during the maximum or minimum time of the solar activity cycles, therefore, it is necessary to study the performance of polar brightening in the transition region. ``The polar layered temporally-staggered brightening" may reflect the different heights of the effects of  magnetic elements of different categories, because brightening phenomena, such as faculae, plages, coronal bright spots, and so on  are closely relates to the magnetic activities.
Therefore studying polar brightening in transition region may help to understand the height of magnetic field effects.
In this study, we will investigate polar brightening  in the transition region through analyzing daily images of the solar full disk observed at the  wavelength of 304 $\AA$.

\section{Data and Methods}
The data used in this research originally  come from  daily images of the solar full disk at the wavelength of 304 $\AA$  during the time interval of 2011 January 1 to 2022 December 31, which were observed by the Atmospheric Imaging Assembly (AIA) on board the Solar Dynamics Observatory (SDO) space mission (Lemen et al. 2012).
Wu et al. (2023) constructed different time series of 304 $\AA$ intensity at different latitudes through scanning one full-disk image observed at almost the same time of each day in the time interval. We will utilize this type of data to investigate  long-term evolution of the transition region, and the data are shown here in Figure 1. The daily sunspot numbers (version 2) during the same interval come from the Sunspot Index and Long-term Solar Observations, the World Data Center, and they are utilized to represent phase variation of the solar activity cycles and displayed here in Figure 2.

\begin{figure*}
\begin{center}
\centerline{\includegraphics[width=1.05 \textwidth]{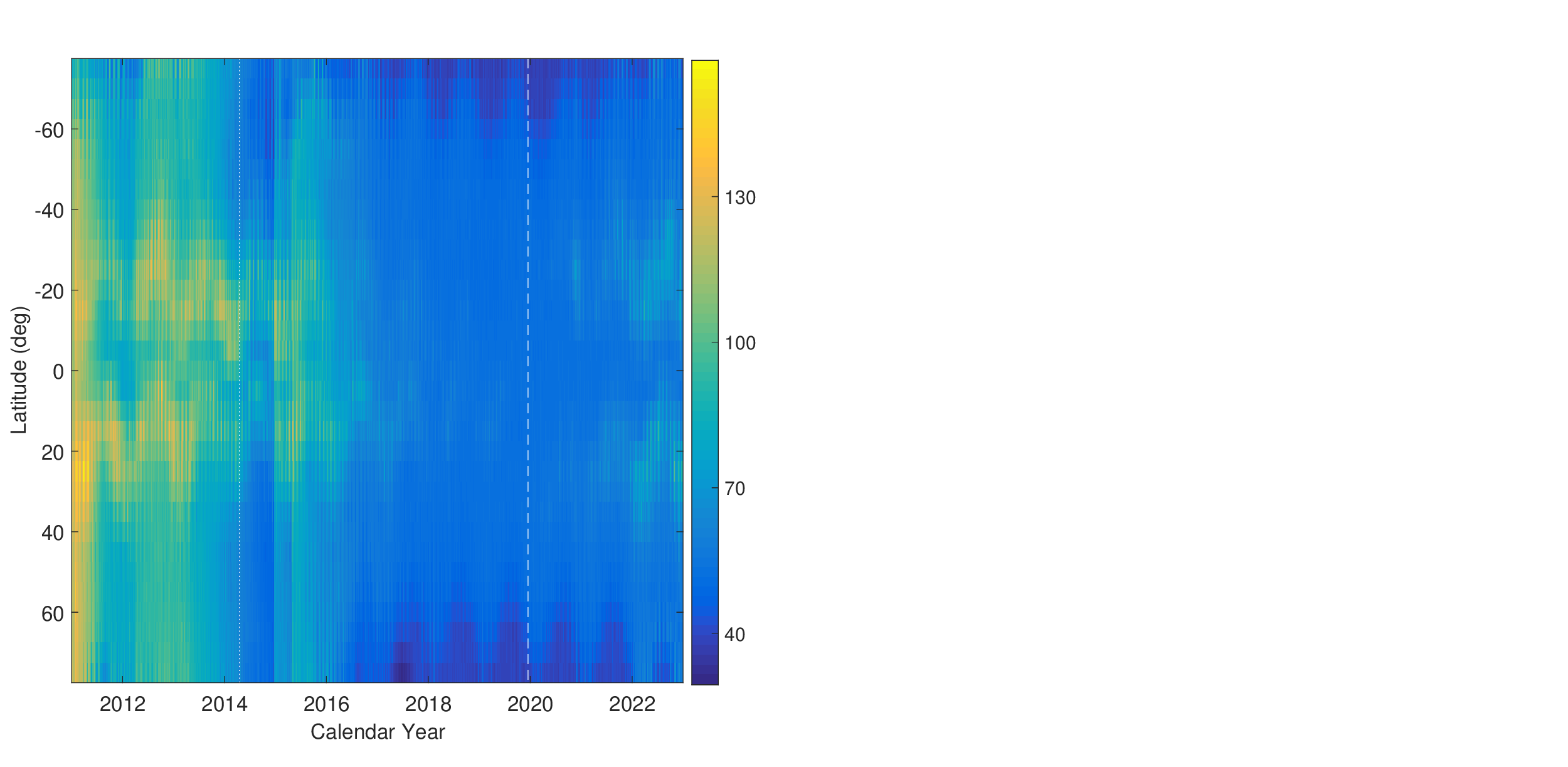}}
\caption{The latitude-temporal distribution of 304 $\AA$  intensity (relative intensity, in arbitrary unit) from 2011 January 1 to 2022 December 31 observed by AIA/SDO. The vertical dashed line indicates the minimum epoch of solar cycle 25.
 }\label{}
\end{center}
\end{figure*}

\begin{figure*}
\begin{center}
\centerline{\includegraphics[width=1.05 \textwidth]{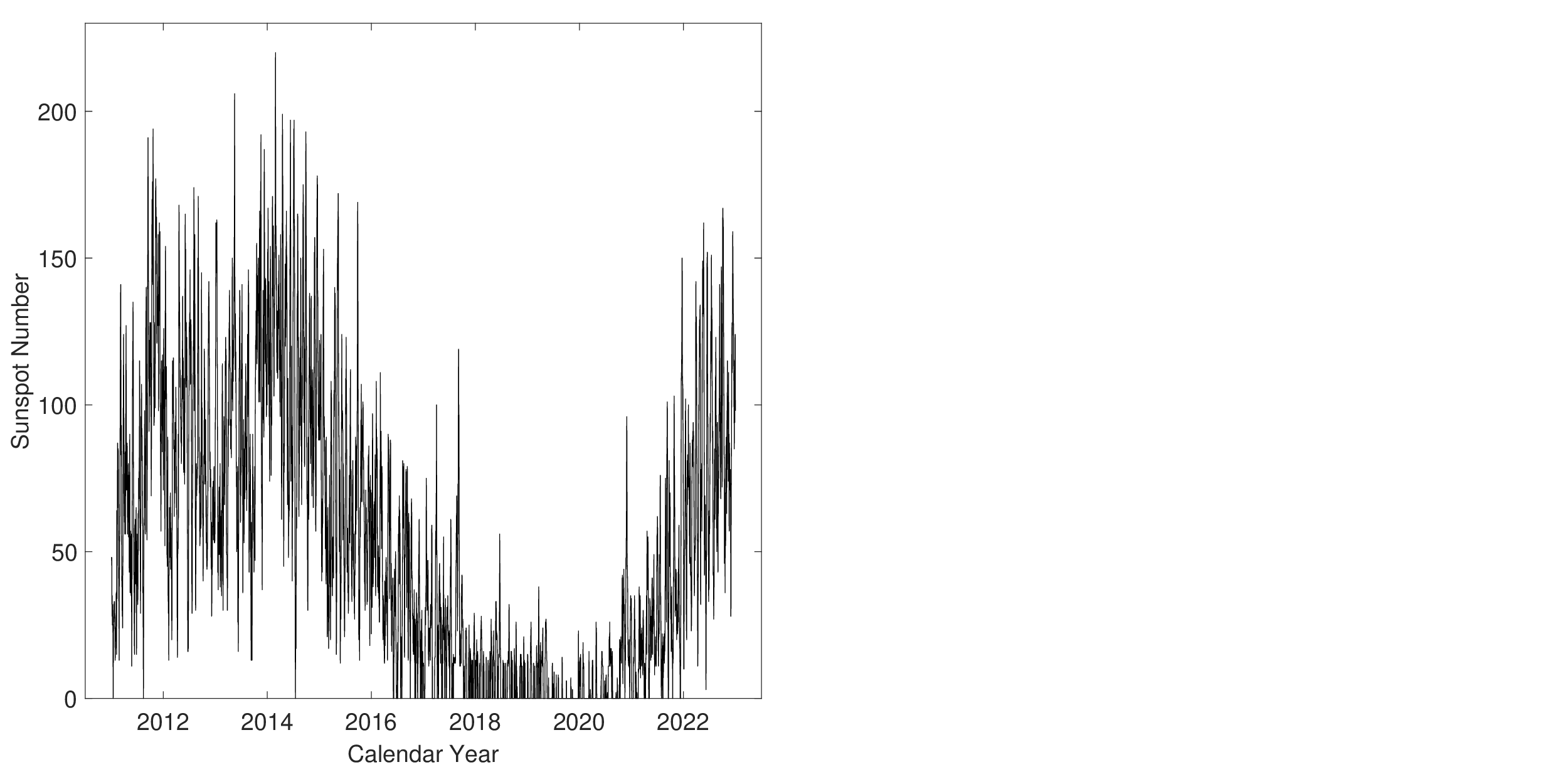}}
\caption{Daily sunspot number (version 2) from 2011 January 1 to 2022 December 31.
 }\label{}
\end{center}
\end{figure*}

As Figure 1 shows, the activity transition region is obviously  characterized by the migration from the latitude of about $35^{\circ}$ toward the equator as sunspot regions do, and its intensity values are generally significantly higher than the background intensity values. Therefore after removing large intensity values, the remaining small intensity values basically correspond to the background quiet intensity ones. Next, we will approach the quiet transition region through eliminating large intensity value as in the previous work (Li et al. 2022a, 2022b, 2024).

\section{Result}
A time series of 304 $\AA$  intensity at each of all latitudes is correlated with daily sunspot numbers, and the obtained correlation coefficient (CC) is shown in Figure 3. The number of 304 $\AA$  intensity used in each CC calculation is 4383, and then the tabulated critical CC value is about 0.06 at the $95\%$ confidence level. As shown in the figure, all CCs are larger than 0.4, and thus the long-term variation of the transition region is significantly positively correlated at the full solar disk with the solar activity cycle, especially at those latitudes where sunspots arise to form the butterfly diagram.

\begin{figure*}
\begin{center}
\centerline{\includegraphics[width=1.05 \textwidth]{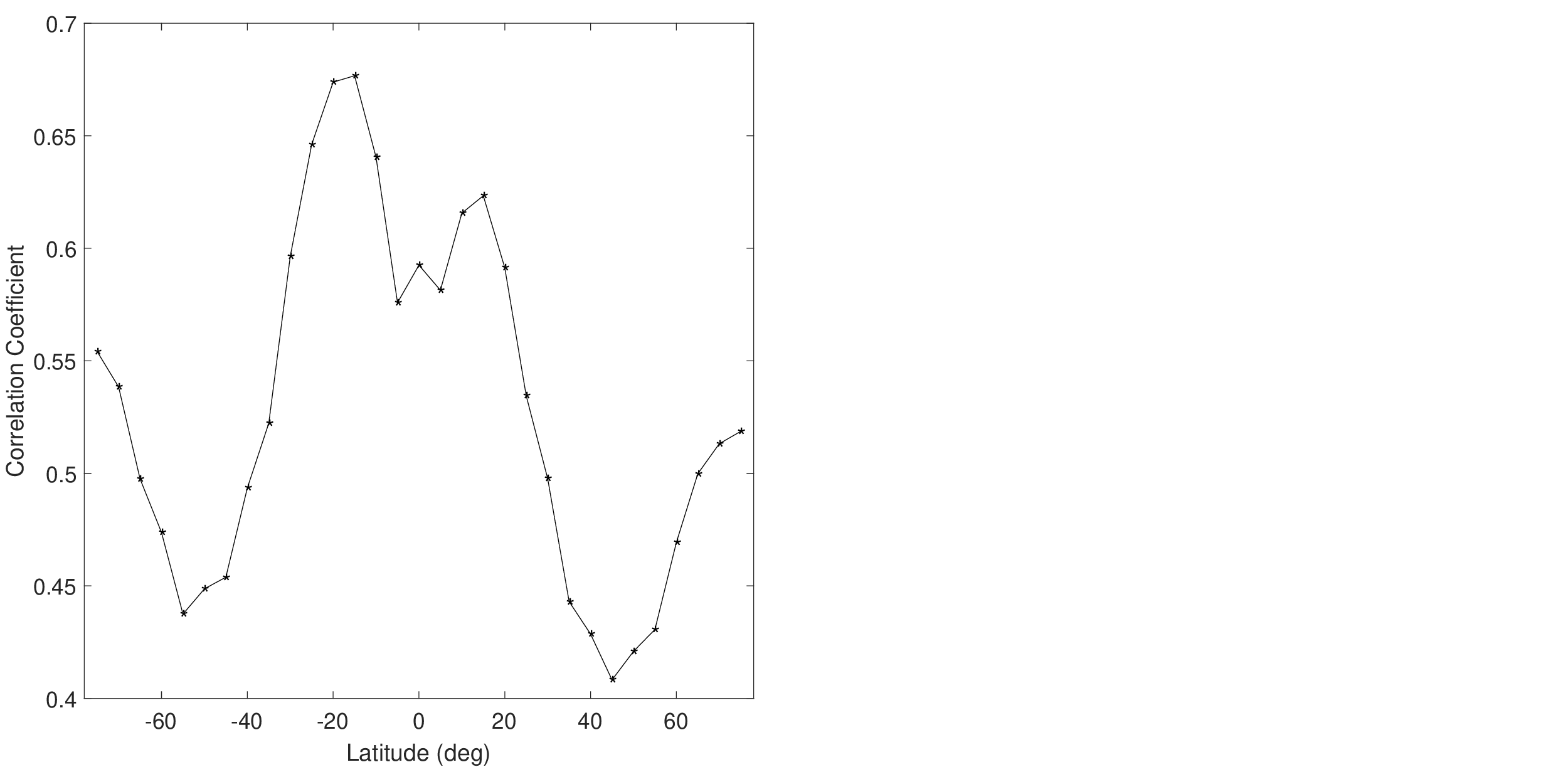}}
\caption{Correlation coefficient (asterisks in the black solid line) of a time series of 304 $\AA$  intensity at a latitude  with  daily sunspot number.
 }\label{}
\end{center}
\end{figure*}

For the entire time series of 304 $\AA$  intensity at a certain latitude, those values which are not greater than $X_{i}$ are selected to perform correlation analysis with the corresponding sunspot numbers, and then a correlation coefficient is obtained. In this case, $X_{i}$ is actually the upper limit value of the selected data from the original time series. We repeat the above process again and again, with $X_ {i}$ decreasing each time, that is, $X_{i}$ decreases continuously from the maximum value of the time series, and finally, we can obtain a series of correlation coefficients changing with  $X_{i}$. Here as an example, Figure 4 shows correlation coefficient at $75^{\circ}$ latitude varying with  the set upper limit value, and correlations at all latitudes are shown in an animation that is attached in the figure.
At all latitudes, correlation coefficient tends to sharply decrease from positive values to negative values, as $X_{i}$ decreases.
At some latitudes, for example at $75^{\circ}$ latitude shown in Figure 4,
as $X_{i}$ decreases, correlation coefficient changes from  statistically significant positive values to statistically significant negative ones.

\begin{figure*}
\begin{center}
\centerline{\includegraphics[width=1.05 \textwidth]{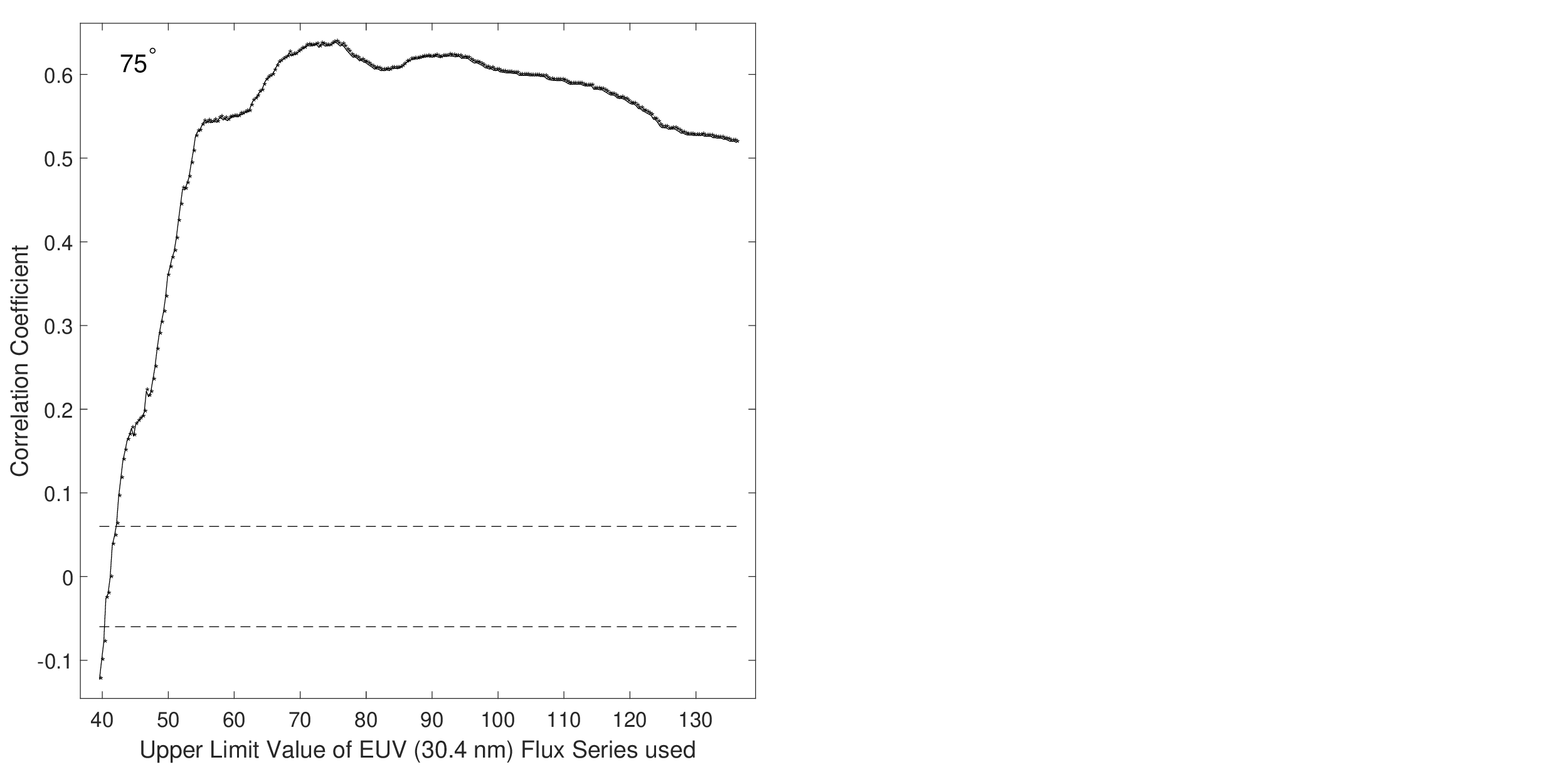}}
\caption{Correlation coefficient (solid points in the black solid line) of the selected 304 $\AA$  intensity values at a certain latitude, for example,  $75^{\circ}$ in this figure (results for all latitudes are shown in the attached animation),  with the corresponding sunspot numbers. Each time, those values which are not greater than $X_{i}$ are selected from the entire original time series of 304 $\AA$  intensity at a certain latitude, that is, $X_{i}$, the horizontal coordinate value of a solid dot, is the upper limit value of the selected data. (An animation of this figure is available.)
 }\label{}
\end{center}
\end{figure*}

Based on the animation, Figure 5 shows the first significantly negative correlation coefficient ($CC_{neg}$) that appears  when  correlation coefficient gradually changes from positive to negative at a latitude with $X_{i}$ decreasing continuously.
As the figure indicates, the long-term variation of background transition region with lower EUV intensity at some latitudes is in anti-phase with the solar activity cycle. Figure 6 displays the number of data used to calculate $CC_{neg}$, namely, the number of data points of the background transition region at a latitude, and that of the entire transition region is 4383. Significantly negative correlation coefficients appear at 14 latitudes, and negative correlation coefficients may emerge at almost all latitudes.

\begin{figure*}
\begin{center}
\centerline{\includegraphics[width=1.05 \textwidth]{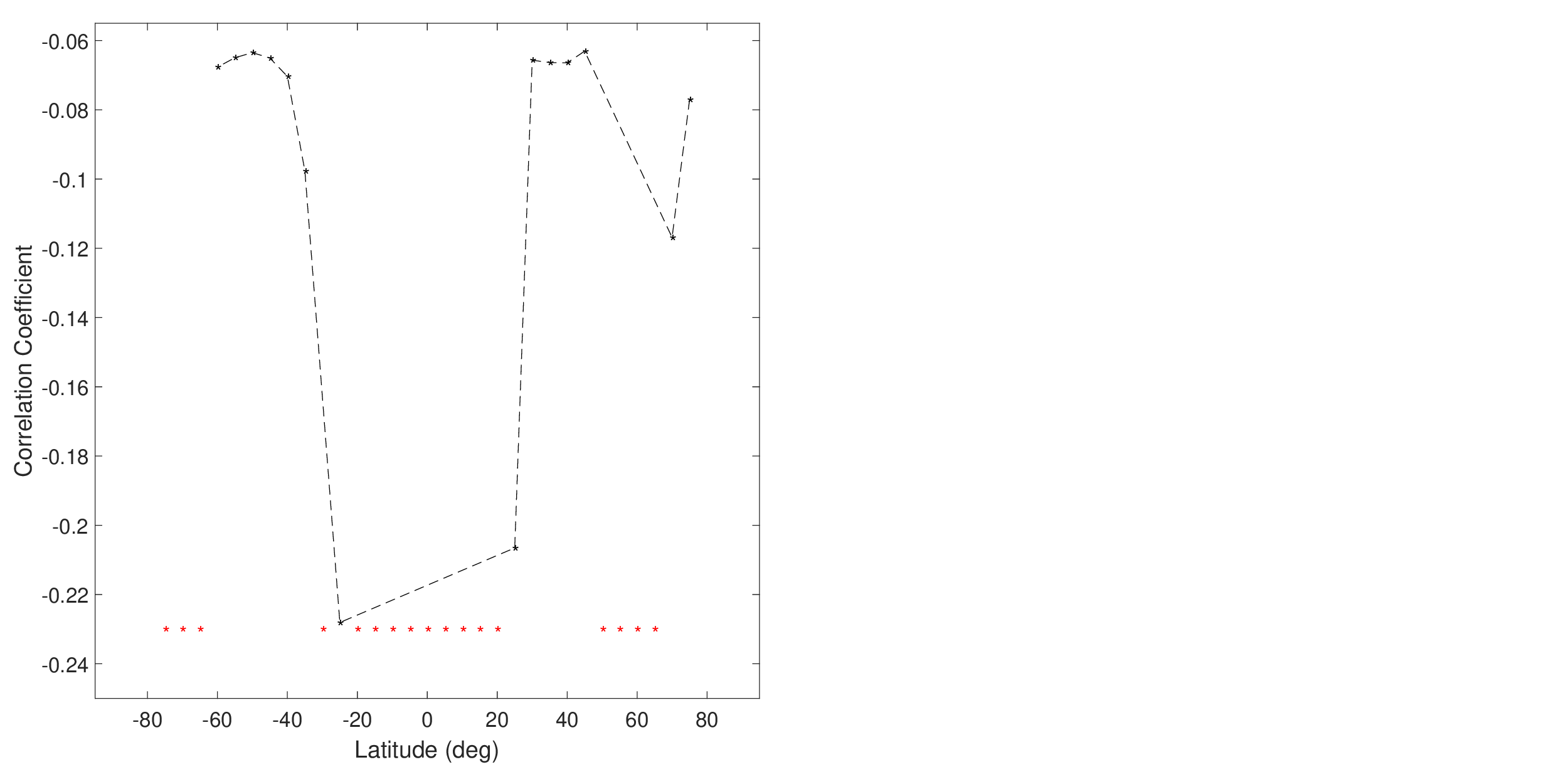}}
\caption{The first significantly negative correlation coefficient (black asterisks in the dashed line) that appears as correlation coefficient gradually changes from positive to negative at a latitude. A red asterisk at a certain latitude indicates that there is no significantly negative correlation to appear at that latitude.
 }\label{}
\end{center}
\end{figure*}

\begin{figure*}
\begin{center}
\centerline{\includegraphics[width=1.05 \textwidth]{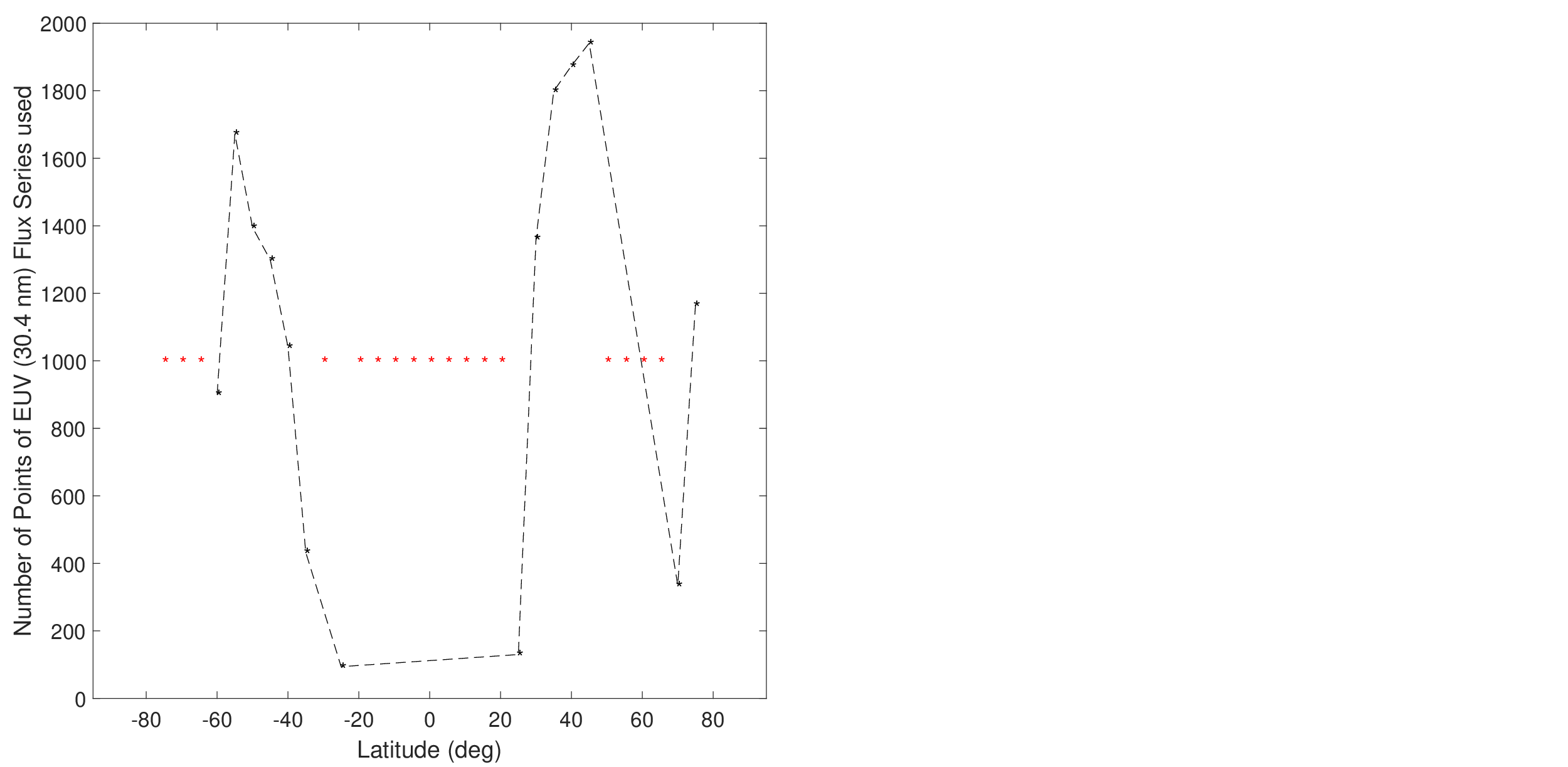}}
\caption{The number (black asterisks in the dashed line) of data used to calculate the first significant negative correlation coefficient at a latitude shown in Figure 5. A red asterisk at a certain latitude indicates that there is not such a significantly negative correlation to appear at that latitude.
 }\label{}
\end{center}
\end{figure*}

As an example, Figure 7 shows  a scatter plot between a series of low 304 $\AA$ intensity and  sunspot numbers at the latitude, $75^{\circ}$. As the figure shows,  data points are diffuse. For the 304 $\AA$ intensity values not larger than 42.2,
the correlation coefficient between the two is -0.077, and it is of statistical significance although this value is small. For the 304 $\AA$ intensity values not larger than 39, the correlation coefficient is -0.176 which is significant, indicating that the correlation improves as the intensity value decreases.

\begin{figure*}
\begin{center}
\centerline{\includegraphics[width=1.05 \textwidth]{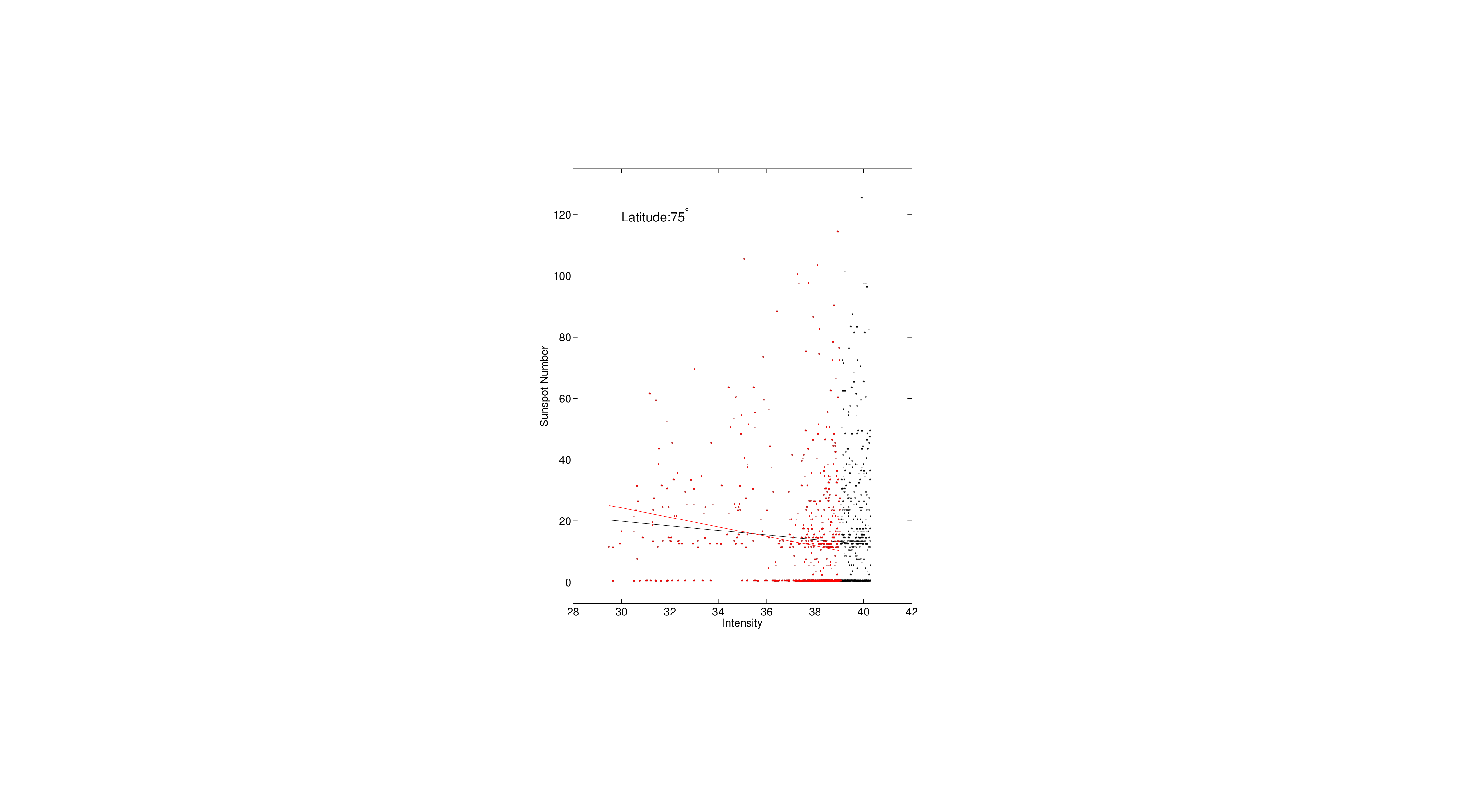}}
\caption{Relationship (red and black asterisks)  between 304 $\AA$ intensity values not larger than 42.2 at the latitude of $75^{\circ}$ and the corresponding sunspot numbers. The black solid line is the linear regression of the red and black asterisks, while the red line is the linear regression of the red asterisks, which are used to show the relationship  between the 304 $\AA$ intensity values not larger than 39  and the corresponding sunspot numbers.
 }\label{}
\end{center}
\end{figure*}

\section{Conclusions and Discussion}
In this study, long-term variation of the transition region is found to be in phase with the solar activity cycle, at all latitudes, i.e. on the entire solar disk, and thus polar brightening should take place in the maximum epoch of the solar cycle.
The complete scene of the ``polar layered temporally-staggered brightening" (Li et al. 2022)  in the entire solar atmosphere should be: in the photosphere and the chromosphere, polar brightening occurs in the minimum epoch of the solar activity cycle, while in the transition region and the corona, it appears in the maximum epoch. The polar brightening in the photosphere and the chromosphere may be caused mainly by the network magnetic fields, for that their long-term variation is in anti-phase with the solar cycle, but the polar brightening in the transition region and the corona may be caused mainly by  the ephemeral magnetic fields, for that their long-term variation is in phase with the solar cycle.

The narrow-band imaging of AIA 304 $\AA$ comes mainly from the emission of  He II at 303.8 $\AA$ with a small amount (about 20$\%$) coming from Si XI at 303.3 $\AA$ (Thompson and Brekke 2000; Lemen et al. 2012). The type of solar features that may be observed is the upper chromosphere and the  transition region (Lemen et al 2012).
The active chromosphere and the active corona are both found to be in phase with the solar cycle, here the active transition region observed by 304 $\AA$ is found to be prominently in phase with the solar cycle as well, therefore the actual active transition region is believed in phase with the solar cycle. The long-term variation of the entire upper solar atmosphere from the chromosphere through the transition region to the corona, which is exactly above the butterfly diagram in the photosphere, is in phase with the solar cycle, and thus the concept ``butterfly body" is proposed to describe such a phenomenon (Li et al. 2022). The upper active atmosphere is inferred to be heated by the magnetic fields of sunspot and the magnetic elements in the ephemeral regions (Li et al. 2022, 2024). Here strangely, the active transition region  doesn't look much like the butterfly diagram to some extent, perhaps due to insufficient spatial resolution of the data used.

After removing the large emission intensity values from the active chromosphere (corona) presented in the butterfly diagram, the remaining small intensity values are considered as the background quiet chromosphere (corona)   (Li et al. 2022, 2024).  Long-term variation of the background quiet chromosphere (corona) is found to in anti-phase with the solar cycle at middle and low latitudes (generally smaller than $\sim 55^{\circ}$), as the network magnetic fields do, and thus the background quiet chromosphere (corona) is believed to be heated by the  network magnetic fields at middle and low latitudes (Li $\&$  Feng 2022a, 2022b; Li et al. 2022).
Here, the transition region is processed using the same method as the chromosphere and the corona to obtain the quiet transition region, and long-term variation of the quiet transition region observed by 304 $\AA$ is found to be possibly in anti-phase with the solar cycle at middle and low latitudes, which is caused also by the network magnetic fields.
The buoyancy experienced by a magnetic element is approximately proportional to the square of its magnetic field strength (Spruit and van Ballegooijen 1982), and thus the ephemeral magnetic fields and the active-region magnetic fields originate from deeper depths within the Sun's interior than the network fields do.
Spatial positions of the network fields are  greatly squeezed out by the appearance of a large number of the ephemeral magnetic fields and the active-region magnetic fields, and thus, sometimes there is no negative correlation between the background transition region and the solar activity cycle at some middle and low latitudes. Insufficient spatial resolution of the used data is believed to affect the measurement of small-scale bright spots, thereby affecting the determination of the relationship between small-scale network magnetic fields and small-scale bright spots.

The ephemeral magnetic fields and the active-region magnetic fields are known to appear mainly at low latitudes (generally smaller than $\sim 35^{\circ}$) as the butterfly diagram (Jin and Wang 2012).
At high latitudes (generally larger than $\sim 55^{\circ}$), the  magnetic fields in ephemeral regions result in a strong positive correlation in the transition region; at the same time, they crowd the network fields to make the anti-correlation in the background transition region less obvious sometimes.
With the increase of latitude, the active-region and ephemeral-region magnetic fields decrease gradually.
At middle and high latitudes (generally larger than $\sim 35^{\circ}$), the magnetic fields in active regions and in ephemeral regions, do not completely squeeze out appearances of the network fields, which makes an anti-correlation occurs in the background transition region.

To sum up, it is believed to be the active-region magnetic fields  and the ephemeral-region magnetic fields that heat the entire transition region, especially the active transition region, so that the heated (active) transition region atmosphere overall  exhibits the same long-term evolution behavior of the solar cycle as these magnetic fields. And it is thought to be the network magnetic fields that heat the background transition region with weak emission at middle and low latitudes, so that the heated background transition region atmosphere  exhibits the same long-term evolution behavior of the solar cycle as these magnetic fields.

At high latitudes, long-term variation of small emission values in the transition region is found to be out of phase with the solar activity cycle, and we do not know whether this reflects variation of the quiet chromosphere or that of the quiet transition region, because the He II line is not a pure optically thin line,
its emission comes from both the chromosphere and the transition region.
Additionally, its formation perhaps slightly influenced by the background corona emission, also affecting the determination of actual long-term variation of the background quiet transition region. Further study is needed in the future.

\acknowledgments
%\begin{acknowledgments}
We thank the anonymous referee very much  for careful reading of the manuscript  and helpful and constructive comments which significantly improved the original version of the manuscript.
The data of  solar full disk 304 $\AA$  images observed by SDO/AIA are courtesy to be
provided by Wu et al. (2023). The daily sunspot numbers (version 2) come from the source: WDC-SILSO, Royal Observatory of Belgium, Brussels, and these can be downloaded from https://www.sidc.be/silso/. The authors would like to express their deep thanks to  these data providers/contributors.
This work is supported by the National Natural Science Foundation of China (12373059, 12373061), the Basic Research Foundation of Yunnan Province (202201AS070042, 202101AT070063), the Yunnan Ten-Thousand Talents Plan (the Yunling-Scholar Project),  the ``Yunnan Revitalization Talent Support Program" Innovation Team Project(202405AS350012), Yunnan Key Laboratory of Solar Physics and Space Science under the number 202205AG070009, the national project for large scale scientific facilities (2019YFA0405001), the project supported by the specialized research fund for state key laboratories, and the Chinese Academy of Sciences.
%\end{acknowledgments}

\end{document}